\documentclass[journal,comsoc,onecolumn,draftcls,10pt]{IEEEtran}

\usepackage[T1]{fontenc}

\ifCLASSINFOpdf

\else
 
\fi

\usepackage{amsmath}

\usepackage[cmintegrals]{newtxmath}
\usepackage{graphicx}

\usepackage[left=1cm,right=4cm]{geometry}

\usepackage{marginnote}
\usepackage{marginnote}
\usepackage{wrapfig}
\usepackage{epstopdf}
\usepackage{cite}
\begin{document}

\title{Non-orthogonal Multiple Access as an Enabler for Massive Connectivity for 5G and Beyond Networks}

\author{Vimal Bhatia, Pragya Swami, Sanjeev Sharma, Rangeet Mitra% <-this % stops a space
\thanks{Vimal Bhatia is with Department of Electrical Engineering, Indian Institute of Technology Indore, India, Email:vbhatia@iiti.ac.in. Pragya Swami is with Department of Electrical Engineering, Indian Institute of Technology Indore, India, Email:phd1601102001@iiti.ac.in.}% <-this % stops a space
\thanks{Received: date / Accepted: date}
}

%{Shell \MakeLowercase{\textit{et al.}}: Bare Demo of IEEEtran.cls for IEEE Communications Society Journals}

\date{}


\maketitle

\begin{abstract}
Two of the most challenging goals to be achieved in the fifth generation (5G) and beyond communication are massive connectivity and higher capacity. The use of traditional orthogonal multiple access techniques limits the number of users that can be served using the available resources due to orthogonality constraint. Moreover, the available resources may not be utilized effectively by the alloted users thereby resulting in inefficiency and user unfairness. This imposes a severe drawback in cases where the number of users to be served are high, like in the 5G networks. Hence, introducing non-orthogonality to the multiple access scheme is advocated as a supreme methodology to serve multiple users simultaneously, thereby enhancing the connectivity. In scenarios with massive number of users, non-orthogonal multiple access scheme increases the number of active connections by superimposing the signal of multi-users on a same resource block, thereby also utilizing the available resources efficiently.
\end{abstract}

% Note thgonal multiple access, 5G communication, Millimeter wave communicaat keywords are not normally used for peerreview papers.

% \fi
\IEEEpeerreviewmaketitle

\section{Introduction}
\label{sec:intro}

Higher capacity and larger number of active users (connected devices) than the current fourth generation (4G) network are the prime goals of the future fifth generation (5G) technology  \cite{KPI_5G}. However, the present use of \textbf{orthogonal multiple access (OMA)} schemes at the physical layer for access methods restricts the number of users that can be simultaneously served using the available resources. OMA, e.g., frequency division multiple access, time division multiple access, code division multiple access, etc.,  works by assigning orthogonal resources to the users, which can not be utilized by other users until the resources are freed by the ongoing user. This can lead to severe degradation in the overall system performance and also limits number of users that can be served at a given instance using the available resource block (RB). This can be explained using a simple example; consider a scenario when a RB is alloted to an user with poor channel gain towards the base station. Due to poor channel, the user can not effectively utilize the alloted RB, neither, another user can use the RB for its own transmission due to orthogonal allocation scheme. This effectively renders the resource waste and hence, it becomes necessary to bring advanced technologies, e.g. new multiple access (MA) scheme to rule out similar possibilities and utilize the available resources efficiently.\leavevmode\normalmarginpar\marginnote{NOMA serves multiple users using the same resources, e.g., frequency, code, etc., at the same instant of time} Breaking the taboo of orthogonality, non-orthogonality has emerged as a leading solution to situations involving wastage of resources, wherein the same RB is shared by multiple users at the same instant of time, hence also increases the number of simultaneous connections. This MA scheme which utilizes the non-orthogonality is termed as \textbf{non-orthogonal multiple acecess (NOMA)}. NOMA serves multiple users using the same resources, e.g., frequency, code, etc., at the same instant of time. One of the prime advantage of using NOMA, by the virtue of its application, is that it can serve a user with better channel condition using the same resource as alloted to the user with poor channel condition, thereby increasing the number of active users and additionally provides efficient utilization of available resources \cite{impactofuserpairing}.

%  One important advantage of the NOMA concept is that it can squeeze a user with better channel conditions into a channel that is occupied by a user with worse channel conditions 
%5G brings out major challenges of higher capacity and larger number of connected users than 4G \cite{KPI_5G}. These requirements are difficult to be satisfied with OMA schemes, which are limited by the number of simultaneously transmitting users and orthogonal resource allocation.To satisfy these requirements, advanced technologies are necessary.

%\newcommand\Right{\textcolor{red}{(Right)As its principle approach, PD-NOMA allocates different power to multiple users with different channel conditions, sends the superimposed signal to the users which then decode their signal using SIC}}

\leavevmode\normalmarginpar\marginnote{As its principle approach, PD-NOMA allocates different power to multiple users with different channel conditions, sends the superimposed signal to the users which then decode their signal using SIC.}Recently, various NOMA schemes have been proposed and discussed. NOMA scheme includes interleave division multiple access \cite{IDMA}, power domain non-orthogonal multiple access (PD-NOMA) \cite{PD_NOMA}, \cite{MUST}, \cite{3GPP_NTT} low density spreading CDMA \cite{low_density_CDMA}, sparse code multiple access (SCMA)\cite{SCMA}, pattern division multiple access \cite{PDMA}, multi-user sharing access \cite{MUSA}. PD-NOMA schemes have been discussed for long term evaluation (LTE) in the third generation partnership programme (3GPP) as multi-user superposition transmission \cite{MUST}. The scheme is approved in the radio access network (RAN) $\#$71 meeting and the discussion of PD-NOMA for new radio (NR) has been initiated since 3GPP RAN1 $\#$84bis meeting in April, 2016 \cite{84BIS}. Hence, this article focuses primarily on the PD-NOMA (referred as NOMA hereafter) scheme and its co-existence with other contemporary and emerging 5G techniques for a wider view of its application in the future generation networks.

\begin{figure}
 \includegraphics[scale=0.45]{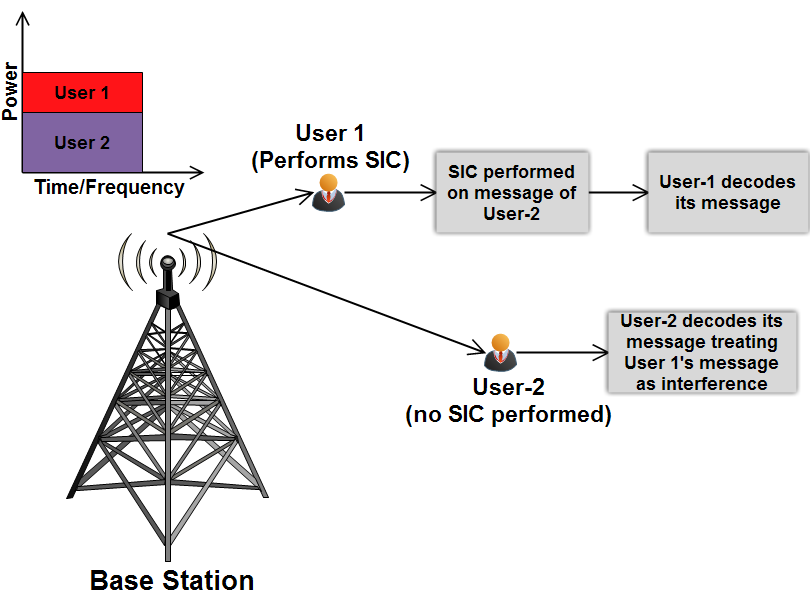}
\caption{ Basic NOMA Concept}
\label{fig:1}       % Give a unique label
\end{figure}

The key idea behind PD-NOMA principle is multiplexing using power diversity. While studying PD-NOMA, three of the foremost terminologies that will be frequently used/referred are: channel difference between users, power splitting and \textbf{successive interference cancellation (SIC)} . As its principle approach, PD-NOMA allocates different power to multiple users with different channel conditions, sends the superimposed signal to the users which then decode their signal using SIC. SIC involves decoding and subtracting/removing message of user with better channel condition such that self-message can be decoded with lower interference from the superimposed signal. For de-multiplexing of the signals at the receiver it is favorable that the difference in allocated power is sufficiently large. This implies that the difference in channel condition between the users multiplexed using NOMA should also be sufficiently large. \leavevmode\normalmarginpar\marginnote{SIC involves decoding and subtracting/removing message of user with better channel condition.}The impact of channel gain difference between the users on the performance gain achieved by NOMA is studied in \cite{VTC}, \cite{GLOBECOM_optimization}. It must be noted that the power allocation in NOMA primarily depends on the channel gain difference between the users and hence plays a vital role in the performance of NOMA systems \cite{GLOBECOM_optimization}. The basic idea of power allocation in NOMA is different from power control, but the algorithm follows similar procedure \cite{saito}.

%PD-NOMA is a multiplexing scheme that utilizes the power domain which is not sufficiently utilized in previous systems. Non-orthogonality is intentionally introduced via power-domain user multiplexing. In fact,user de-multiplexing is ensured via the allocation of large power difference between paired users and the application of SIC in power-domain. It is different from the simple power control, but power distribution by the base station is conducted according to the related algorithm \cite{saito}.
Also, different from the conventional power splitting techniques which allots more power to users with better channel condition, in PD-NOMA the power allocation is inversely proportional to the channel condition of the user. This implies that a weak user, i.e. a user with poor channel condition will be awarded larger power. This balances the trade-off between system throughput and user fairness which imposes prime importance in any wireless communication system. The traditional power splitting technique achieve better overall system performance, however, at the cost of weak user's throughput. Also, it is important to note that although the power splitting in NOMA reduces the power given to a particular user as given in OMA, however, the achieved performance gain is due to the assigned bandwidth which otherwise is distributed amongst the users served using OMA. Furthermore, the factor that makes NOMA even more popular is its compatibility with OMA receivers and transmitters which require only slight modifications in the software and little or no hardware changes for its implementation. Hence, NOMA is advocated as an ideal enabler for massive connectivity in 5G networks and beyond.

%But unlike these existing techniques,NOMA seeks to strike a balance between throughput and fairness. For example, the transmission power allocated to the users in NOMA systems is inversely proportional to their channel conditions, which is important to ensure that all the users are served simultaneously. On the other hand, conventional opportunistic schemes prefer to give more power to users with better channel conditions, which can improve the overall system throughput but deteriorate fairness.
%In NOMA,although user multiplexing through power allocation reduces the allocated power to each user, users with higher channel gains (NOMA strong user) and users with smaller channel gains(NOMA-weak user) benefit from being scheduled more frequently with more assigned bandwidth. Moreover, NOMA is compatible with orthogonal frequency division multiple access in the downlink (DL), thus many users can be served simultaneously, making NOMA ideal for massive connectivity and low latency transmission in 5G networks.

\section{NOMA and its Application to Three Main Use-Cases of 5G}\label{sec:5Gapplication}

The 5G technology is categorized as three basic use-case scenarios namely enhanced mobile broadband (eMBB), massive machine type communications (mMTC), and ultra-reliable and low latency communications (URLLC) \cite{recommendation_ITU_5G}. \leavevmode\normalmarginpar\marginnote{The 5G technology is categorized as three basic use-case scenarios namely eMBB, mMTC and URLLC.}Since, NOMA is envisioned as a prime MA scheme for the 5G technology it becomes important to discuss its application for the practical use cases of 5G technology. The expected purposes of advanced MA scheme for 5G are: achieving multi-user capacity boundary, aid overloaded transmissions, and avail low latency grant free transmission. All such functionalities are fulfilled by NOMA, hence renders itself suitable for mMTC application which requires massive connectivity. Moreover, the benefits of NOMA comes with additional receiver complexity, however, the mMTC is expected to be uplink centric. In the uplink communication, the user equipments (UE) are not prone to receiver complexity, rather the complexity increases at the network side which is not a big issue. Furthermore, mMTC requires only small packet transmissions. For small packet transmission in uplink scenario, the UEs do not require explicit need of scheduling grant from the evolved NodeB in order to reduce the power consumption at the battery operated UE side and also for lowering the overhead and latency. NOMA has proven to yield better performance than its counterpart MA scheme, OMA, in the grant-free transmission. This is because, bandwidth available for the users is larger, thereby reducing the collision probability when the same resource is used by multiple users. It is also notable that such small packet transmission are not only a part of mMTC transmissions. Similar traffic is also present in the eMBB and URLLC \cite{multipleaccessforUL}. As per the discussion on the grant-free uplink transmission, NOMA is rendered suitable also for the eMBB and URLCC application. For above basic usage scenarios of eMBB, mMTC and URLLC, typical use cases like relay and V2V are analyzed in \cite{usecases_5G}.

\section{NOMA with Contemporary Emerging Technologies}\label{sec:combination}
\begin{figure*}
 \includegraphics[scale=0.4]{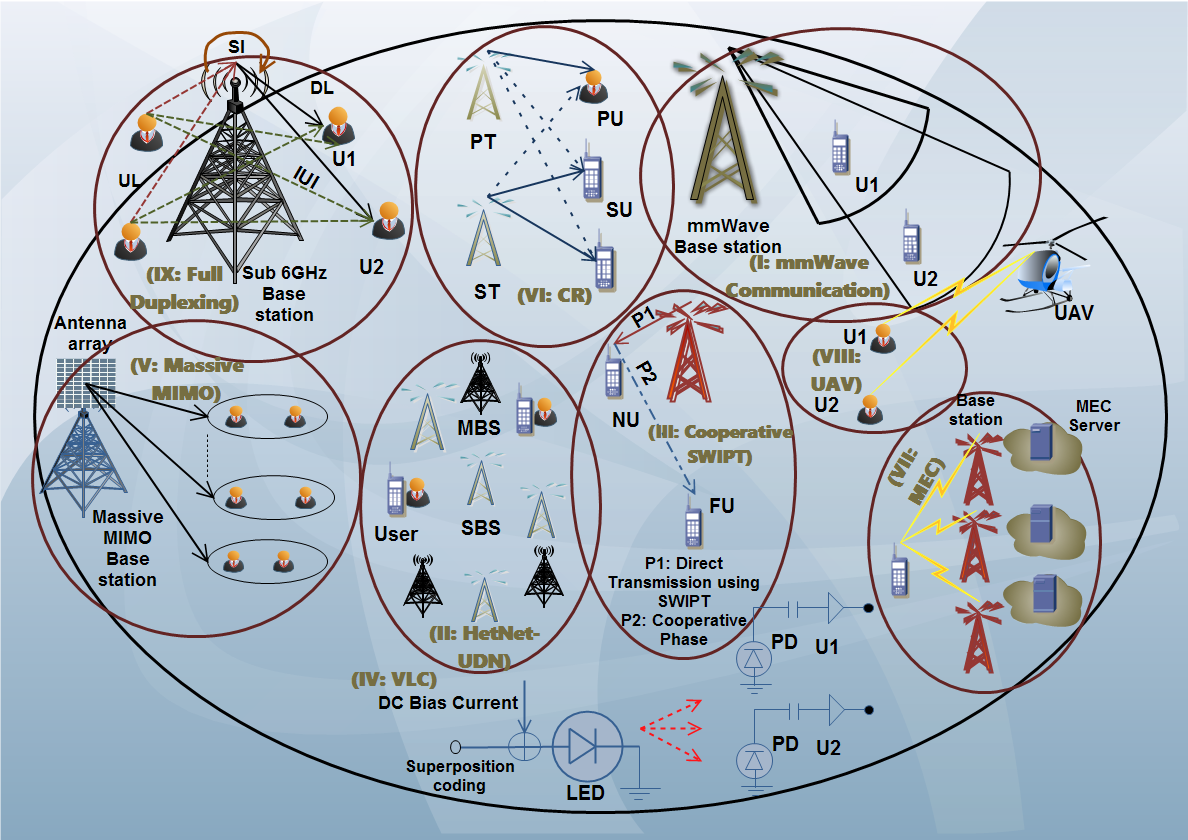}
\caption{NOMA with various emerging technologies for 5G technology}
\label{fig:2}       % Give a unique label
\end{figure*}

Various other emerging technologies contemporary to NOMA for achieving the goals of 5G technologies are multiple-input multiple-output (MIMO) and massive MIMO, millimeter wave (mmWave) communication, cognitive radio (CR), cooperative communications, energy harvesting (also termed as simultaneous wireless information and power transfer (SWIPT)), visible light communications (VLC), mobile edge computing (MEC), etc. Few other technologies that have recently come out and have been studied for application in the future generation network are unmanned aerial vehicle (UAV), full duplexing (FD) and ultra dense networks (UDN), to name a few. The key idea of any developing technology is to enhance the system performance in the given available resources and serve as many number of users as possible. To utilize the available resources, MA schemes are vital to be studied in context of all the upcoming technologies.  Owing to the flexibility of NOMA in its implementation, it can be easily integrated with most of the other fellow techniques. Hence, this articles explores various applications of NOMA with the available emerging technologies to meet the diverse requirements of 5G and beyond cellular network as shown in Fig.~\ref{fig:2}. 
%NOMA being the most trending MA scheme for the future networks, it become of prime interest to discuss the application of NOMA in the contemporary  counterparts.

%In order to fulfill the diverse requirements of 5G and beyond cellular networks, several new technologies have been developed during the past decade. Among them is non-orthogonal multiple access (NOMA) [1]–[3], which can help to address the above challenges more efficiently than the conventional orthogonal multiple access (OMA) schemes. NOMA can be flexibly combined with many other existing and emerging technologies, such as multiple-input multiple-output (MIMO)and massive MIMO, millimeter wave, cognitive and cooperative communications, physical layer security, visible light communications, energy harvesting, mobile edge computing,etc. NOMA can be combined with these technologies to further increase the number of users and enhance the system performance in various senses.

\subsection{NOMA with mmWave Communication}\label{sec:mmWave}

Large\leavevmode\normalmarginpar\marginnote{Large chunks of underutilized bandwidth are available at higher frequencies referred to as mmWave frequency, that can resolve congestion and also provide higher bandwidth as compared to the limited resources available for the 4G communication.}
chunks of underutilized bandwidth are available at higher frequencies referred to as \textbf{mmWave frequency}, that can resolve congestion and also provide higher bandwidth as compared to the limited resources available for the 4G communication \cite{chandra_mag}. Although the accessible bandwidth is huge at the mmWave frequencies, NOMA can be utilized as an essential MA technique. The physical nature of the mmWave communication makes NOMA even more preferable for its environment, as shown in Fig.~\ref{fig:2} (I). The mmWave communication is very directional in nature owing to its large path loss which requires suitable antennas to focus the waveform in a particular direction to cope with the high losses while propagation. The directional property makes the channel gain of users correlated. Such correlation leads to severe degradation in the conventional OMA technique, however, for NOMA application, channel correlation results in boosting the overall performance \cite{NOMAmmW7}. The wide bandwidth available at the mmWave frequency are capable of supporting large number of users, but the sparse channels in mmWave communication limits simultaneous connections of multiple users \cite{chandra_mag}, \cite{mmW_outdoor}. Hence, NOMA with mmWave communication can achieve tremendous increase in connectivity especially in the overcrowded areas. Furthermore, spectral efficiency is vital for the mmWave communication with the upcoming growth of virtual and augmented reality which is said to lower the radio spectrum gains obtained by the mmWave bands. Furthermore, standalone mmWave base stations are not a feasible choice for a practical system. The reason being the delay caused in the initial access \cite{initial_beamtraining} using thin beams (process known as beamtraining). The literature suggests the future wireless network will comprise a hybrid  network with Sub-6 GHz base stations co-existing with mmWave base stations. Such hybrid network can rule out the high access delay in beamtraining \cite{RAT}, \cite{ISWCS}.

%
%Even though more bandwidth
%resources are available at very high frequencies, the use of
%NOMA is still important for the following reasons:
%
%
%• The highly directional nature of mmWave transmission
%implies that users’ channels can be highly correlated,
%which potentially degrades the system performance. But
%such correlation is ideal for the application of NOMA.
%• The combination of NOMA and mmWave supports massive connectivity in dense networks, e.g., where there are
%hundreds of users to be connected in a small area.
%• The rapid growth of mobile Internet services, particularly emerging virtual reality (VR) and augmented reality (AR) services, will dwarf the radio spectrum gains
%obtained from the mmWave bands, which means that further improvement of spectral efficiency is still important.
%\subsection{NOMA with Energy harvesting}

%since mmWave
%channels are sparse in spatial/angle domain, the number of
%simultaneous connections at these very high frequencies has
%shown to be limited. Coupled with mmWave massive MIMO,
%NOMA can circumvent this limit
\subsection{NOMA with Heterogeneous Networks and UDNs}\label{sec:UDN}

\textbf{UDN} refers to a wireless environment where the number of access points (APs) are larger than the number of active users. Large number of base stations are deployed for extensive spatial reuse, another way to utilize spectrum efficiently \cite{UDN_mag}, \cite{UDN_NOMA_Mag}. 
\leavevmode\normalmarginpar\marginnote{UDN refers to a wireless enviroment where the number of access points are larger then the number of active users.}Future generation networks are moving towards \textbf{heterogeneous networks (HetNets)}, which refers to a network where different types of base stations co-exists to meet the requirements. To fulfill different requirements, macro base stations (MBS) and small base stations (SBS) are deployed in same space. The MBSs are capable of providing the required wide coverage while the SBSs boosts the capacity of the network and enhances the data rates due to small coverage areas. NOMA integrated with such HetNets has been extensively studied in the literature as in \cite{MY_TVT}, \cite{MY_ISJ}. Furthermore, heterogeneous UDNs require that the deployment of the MBSs and the SBSs follow the rule set as per the UDN network demands, i.e., number of base stations needs to be larger than the number of users, as shown in Fig.\ref{fig:1} (II). This implies that in a heterogeneous UDN the MBS and SBS will have a very dense deployments. Although,  \leavevmode\normalmarginpar\marginnote{Heterogeneous networks refers to a network where different types of base stations co-exists to meet the requirements.} network densification improves the capacity, however, there exists a limit posed due to the aggregate interference cause by the interfering base stations. Denser the deployment of the base stations, more is the interference suffered by neighboring users \cite{network_densification}. However, towards a brighter side, dense deployments ensure that each base station serves fewer users. Moreover, the distance between the users and base stations becomes less, thereby lowering the impact of path loss. To null the dominant impact of the increased interference of the dense networks, several innovative technologies are studied to design the architecture of heterogeneous UDNs to make the management of the network easy \cite{architecture_5G}. Several technologies used are software-defined networking (SDN) \cite{SDN}, network function virtualization (NFV) \cite{SDN}, cloud computing \cite{cloud_computing}, and fog computing \cite{fog_computing}. Implementation of UDNs implies massive number of connections with high data rates. The access points are connected to a server for the control and management of user access. Applying NOMA protocols where signals of different uses are multiplexed over a single RB, the number of active connection can be further boosted. In context of UDNs, where the number of users and APs are nearly similar, NOMA can be implemented by multiplexing different APs over the same RB \cite{UDN_mag}, \cite{UDN_NOMA_Mag}.

However, integrating NOMA in UDN poses another challenge of severe inter cell interference from both intra tier and inter tier BSs and intra cell or inter user interference (IUI) caused by NOMA which require thorough study of the inter cell-interference management.

\subsection{NOMA with Cooperative Communications using SWIPT}\label{sec:SWIPT}

The coverage for the cell edge users, i.e., users located at the cell edge away from the base stations, can be improved by using \textbf{cooperative communication} using relays which receive the signal from the base station and retransmit it to the cell edge users. 
\leavevmode\normalmarginpar\marginnote{Using a cooperative communication reduces the attenuation suffered while the signal propagates to reach cell edge user.}Several types of relaying techniques have been studied in the existing literature \cite{AF_DF_CR}, \cite{AF_DF_NOMA}. Combining NOMA with cooperative communication is a natural extension, since as per its principle, NOMA receives a superimposed signal of multiple users multiplexed using NOMA. The cell center user, i.e., the user located near the cell center, applies SIC before decoding its self-message. This implies that the cell center user has the message of the cell edge user, i.e., the user located at the cell edge. Hence, the signal of cell edge user can be retransmitted by the cell edge user, for e.g., in scenarios when direct link does not exists between the base station and cell edge user. The use of cooperative communication reduces the attenuation suffered while the signal propagates to reach the cell edge user \cite{full_duplex_cooperative_NOMA}, \cite{ICACCI}, \cite{IJCS}, since the signal is retransmitted from a node/user located closer to the cell edge user. This enhances the performance of the cell edge user. However, it must be noted that the UEs or the relays are generally battery operated device and retransmitting the signal can drain their batteries quickly. Hence, some advanced approach is required for an energy efficient design. In this context, the concept of energy harvesting (known as SWIPT in wireless communication) strives to be of great interest for the researchers. The concept of \textbf{SWIPT}, first proposed in \cite{SWIPT_Firstpaper}, have gained attention of many to design a energy efficient networks. NOMA applied with SWIPT for cooperative communication will not only lead to spectral efficiency but can yield an energy efficient approach for serving the cell edge user using relaying/cooperation. The authors in \cite{SWIPT_Firstpaper} assumed that using SWIPT technology, the radio signals can be used in a two-fold manner, first use is for the information transfer and the second use is for harvesting of energy.\leavevmode\normalmarginpar\marginnote{Using SWIPT technology, radio signals can be used in a two-fold manner, first is for the information transfer and second is for harvesting energy.} However, one of the assumption that was not practical in \cite{SWIPT_Firstpaper} was that both the information transfer and energy extraction was assumed to be carried out at the same instant of time. A more practical and efficient way for SWIPT is to split it into two distinct tasks which are performed in two phases, one for energy extraction and second for information transmission. The energy harvested in the first phase by the relays/cooperating users can be used in the second phase for information transfer to the cell edge user which is a very energy efficient approach. Also, as discussed in Section~\ref{sec:intro}, in NOMA, a cell center user is served using the same resource as that allotted to the cell edge user. Hence, it is straightforward to assume that the performance of cell edge user will deteriorate due to power splitting and interference from the cell center user. Hence, using the cell center user to assist in relaying the information to the cell edge user, as shown in Fig.~\ref{fig:2} (III), compensates such sharing of resources and also increases the spectral efficiency of the system \cite{cooperative_NOMA_Ding}. Furthermore, using SWIPT enabled cell center users rules out the possibility of battery drainage during relaying.

\subsection{NOMA with VLC}
Beyond \marginnote{VLC is a shift towards the nano-meter wave range where the available bandwidth can be expanded so as to meet the spectrum demands for the future generation networks.}the radio frequency communication lies yet another branch of communication that can act as its supplement and is commonly known as \textbf{VLC} \cite{VLC_Mag}. VLC  is a shift towards the nano-meter wave range where the available bandwidth can be expanded so as to meet the spectrum demands for the future generation networks. VLS is a core part for the Light Fidelity (LiFi) systems which is a proposed technology for various applications, for instance, Internet of Things devices, underwater communications, vehicle-to-vehicle communication, etc. VLC comprises of light emitting diodes (LED) lamps, which generally are used a a source of light, as optical transmitters, as shown in Fig.\ref{fig:1} (IV). At the receiver end, photodiode-arrays are used in a typical optical attocell \cite{NOMA_VLC}. The modulation by the LEDs is done at a speed that can not be perceived by the human eye \cite{LED_nonlinearity}, hence making the use of LEDs two-fold, without one interrupting the other. One use is for illumination and second one for communication. Moreover, the power required by the LEDs for signal transmission is very low as compared to the equipments used in radio frequency communication. NOMA is proposed as a viable option for candidate MA scheme in VLC. The major reasons for considering NOMA potential to be used in VLC are \cite{NOMA_VLC2}: VLC poses a very small communication range, hence the channel is dominated by the line-of-sight (LOS) component which promotes accurate channel estimation. \marginnote{VLC comprises of LED lamps, which generally are used a a source of light, as optical transmitters. At the receiver end, photodiode-arrays are used in a typical optical attocell}Channel estimation quality plays a vital role in the performance of the NOMA systems, since the SIC carried out at the receiver which involves decoding and removal of the messages until desired signal yields. SIC depends on the channel estimation of users. In case of imperfect SIC, the error propagates at every step of SIC and the accumulated error becomes so high that the overall performance degrades. Secondly, in VLC, the two parameters other than the channel gain, namely, tuning angle and field of view, gives additional degree of freedom, which can be utilized for the multiplexing multiple users' signals. 
Although current literature has investigated the integration of NOMA in VLC, however, the gains expected are limited due to the non-linearity posed by the LEDs \cite{double_heterostructure_LED}. As a solution, the non-linearity inherited by the LEDs can be eliminated using pre-distortion or post-distortion techniques. The authors in \cite{mitra} takes into consideration the non-linearity posed by the LEDs to design a pre-distorter. For this, a modified Chebyshev-NLMS based pre-distortion is proposed with hybrid eigen-decomposition based precoding in a MIMO NOMA-VLC setup.

\subsection{NOMA with Massive MIMO}\label{MIMO}

\textbf{MIMO} \leavevmode\normalmarginpar\marginnote{MIMO refers to use of multiple antennas at the transmitter and receiver so that multi-path propagation can be exploited.}
refers to use of multiple antennas at the base station for transmission and using multiple antennas at the receiver end so that multi-path propagation can be exploited to increase the system performance  \cite{Massive_mimo_mag}. MIMO has emerged as a key component and has been included in the new LTE-Advanced and 5G deployments \cite{MIMO_mobile_broadband}. From a theoretical concept assuming unlimited number of antennas \cite{unlimitedantennas} to practical implementation which now has been deployed in the LTE-Advanced network uses $64$ antennas at the base station, MIMO has gained tremendous popularity. Massive-MIMO refers to a scenario where the base stations are equipped with large number of antennas as compared to the number of users served by the base station. Application of NOMA in MIMO networks, as shown in Fig.~\ref{fig:2} (V), have been well analyzed in \cite{NOMA_applcation_to_MIMO}, \cite{MIMO_NOMA}. However, \normalmarginpar\marginnote{Massive-MIMO refers to a scenario where the base stations are equipped with large number of antennas as compared to the number of users served by the base station.}not many works exists which analyses the impact of NOMA when applied to massive MIMO setups. However, one of the recent literature has discussed the role of applying NOMA to environment with multi-antenna base stations for performance enhancements \cite{NOMAroleinMassiveMIMO}. The study compares the NOMA-MIMO setup with the conventions MIMO setup with a typical zero-forcing (ZF) massive MIMO beamforming technique and suggest that the NOMA does not outperforms ZF scheme in every scenario. However, in situation where the number of users are very large, NOMA has an edge over ZF scheme. Hence, proving the role of NOMA for massive connectivity.

%Another key technology is massive MIMO,
%which refers to the use of base stations (BSs) with a large number of antennas \cite{Massive_mimo_mag}. we investigate the application of NOMA
%at BSs that are equipped with many antennas, since that appears
%to be the norm in the new LTE-Advanced and 5G deployments
%\cite{MIMO_mobile_broadband}. 
%In less than a decade, massive MIMO has transitioned
%from being a far-fetched theoretical concept with an unlimited number of antennas \cite{unlimitedantennas} to a practical technology that has
%been commercially deployed in LTE-Advanced networks using
%64-antenna BSs.
% In a nutshell, massive MIMO refers to systems where the BSs are equipped with a large number of
%antennas, M, as compared to the number of simultaneously active users, K. NOMA can also be applied to multi-antenna networks for performance improvement \cite{MIMO_NOMA}.

\subsection{NOMA with Cognitive Radio Networks}\label{sec:cognitive}

\textbf{CR} network is one of the most spectral efficient approach for sharing the spectrum available for communication. The CR networks use adaptive intelligent method that detects the availability of vacant spectrum (called as spectrum holes) and enable concurrent transmissions to serve the secondary users (SU) in the spectrum holes. \leavevmode\normalmarginpar\marginnote{The CR networks uses adaptive intelligent method that detects the availability of vacant spectrum (called as spectrum holes) and enable concurrent transmissions in the spectrum holes.} Various techniques like underlay, overlay, and hybrid have been proposed which decides the manner in which the spectrum sharing takes place \cite{underlay_overlay}. In the underlay approach, secondary transmissions (by the secondary transmitter (ST)) takes places together with the primary transmission (by the primary transmitter (PT)) conditioned on the limit of the interference caused at the primary user (PU). Using the underlay technique, transmission takes place only when spectrum holes are detected else no transmission is performed. Hybrid approach uses an intelligent combination of underlay and overlay techniques as per the system requirement.
Although researchers have been looking into integrating CR networks with NOMA, as shown in Fig.~\ref{fig:2} (VI), as one of the possible and effective combination for the 5G technology, the interference caused in the CR networks due to NOMA poses a challenging difficulty to be tackled. The CR-NOMA networks are envisioned to be spectrally efficient with high system capacity. Some of the existing literature have investigated the performance of NOMA integrated with CR networks as in \cite{NOMA_CR1}, \cite{NOMA_CR2}. CR networks combined with NOMA can effectively increase the number of users served over an RB.
%
%adaptive, intelligent radio and network technology that can automatically detect available channels in a wireless spectrum and change transmission parameters enabling more communications to run concurrently and also improve radio operating behavior

%CR and NOMA
%are envisioned to be important solutions for fifth
%generation wireless networks. Integrating NOMA
%techniques into CRNs has tremendous potential
%to improve spectral efficiency and increase system capacity. However, there are many technical
%challenges due to the severe interference caused
%by using NOMA. Many efforts have been made
%to facilitate the application of NOMA into CRNs
%and to investigate the performance of CRNs
%with NOMA.

\subsection{NOMA with MEC}\label{sec:MEC}

\textbf{MEC} is capable of meeting the requirement for real-time mission-critical application such as vehicular networks, connected cars, health care, smart venues, edge video caching, and many others. MEC extends the centralized cloud computing capability to the UEs/devices located near to the edge. \leavevmode\normalmarginpar\marginnote{The key idea behind the use of MEC is that the application performance is much better when the related processing and tasks runs closer to the UE since it lowers the network congestion.} The key idea behind the use of MEC is that the application performance is much better when the related processing and tasks runs closer to the UE since it lowers the network congestion. MEC can utilize the mobile backhaul quite efficiently, however there still exists several challenges, such as, ultra-low latency, energy-efficient computation, etc. The 
latest inventions like virtual reality, augmented reality, interactive gaming renders the mobile networks constrained by virtue of computations. For instance, virtual reality requires UEs to handle several tasks such as object recognition, vision base tracking, etc. For gaming based on virtual reality, the UE will have to carry out mixed reality and human computer interaction \cite{virtual_reality}. However, the major issues with the UEs is that they are not so efficient in terms of handling high computational complexity tasks and that they are battery operated, hence have limited power source. Intensive computations may lead to complete drainage of batteries of UEs which further results in incomplete deadlines. Therefore, the MEC operate by providing and deploying facilities for computations at the edge of the network. Such facilities involve APs/SBSs with MEC servers, as shown in Fig.~\ref{fig:2} (VII). The UEs can offload their tasks to the MEC facilities provided. The collaboration of NOMA and MEC is seen as another important communication technique for future generation wireless networks. The benefits of such collaborations has been studied in \cite{NOMA_MEC1} and \cite{NOMA_MEC2}.

\subsection{NOMA-aided UAV networks}\label{sec:UAV}

\textbf{UAVs}\leavevmode \normalmarginpar\marginnote{UAVs or drones are aircrafts with no human on-board to control their movements.} or drones are aircrafts with no human on-board to control their movements. The movements of UAVs are managed using a ground-based controller and a system of communication is setup between the ground-based controller and the UAV. The initial use of UAVs were meant only for military purposes, however, with time the potential of UAVs for civil purposes emerged as a possible topic of discussion amongst the researchers with the growth of technology and manufacturing. Various applications of UAVs in the civil scenarios include wildfire management, distribution of cargos, aerial photography, etc. \cite{airborne_comm}. Both industries and academia are showing keen interest in recognizing the implementation of UAVs in the field of communication. For instance, some of the industrial application include Google Loon project, Facebook's delivery drone \cite{facebook_drone}, and airborne LTE services from AT$\&$T for availing global massive connectivity. In order to integrate the UAVs for the communication purposes in the 5G networks, it becomes essential to explore suitable MA schemes. Since, NOMA is already a popular MA scheme for the 5G networks, combination of UAVs with NOMA, as shown in Fig.~\ref{fig:2} (VIII) is in line as a promising candidate for future wireless connectivities \cite{UAV_NOMA}.

%With the rapid development of control technology and manufacturing business, unmanned aerial vehicles (UAVs), which were originally sparked by military use, have gradually demonstrated the civil potentials for new applications and market opportunities, such as advanced cargo distribution,aerial photography, and wildfire management, to name a few \cite{airborne_comm}. Regarding communication areas,UAV-aided communication has been recognizedas an emerging technique by both industry andacademia for its superior flexibility and autonomy. On one hand, industry projects, such as the Google Loon project, the Internet delivery drone of Facebook \cite{facebook_drone}, and airborne LTE services fromAT$\&$T, have been deployed for providing airborne global massive connectivity. On the road to integrating UAV into 5G networks and beyond, multiple access techniques are essential. Currently, non-orthogonal multiple access(NOMA) is regarded as a promising candidate for 5G networks due to its superiority in providing higher spectral efficiency and supporting massive connectivity therefore integrating the two techniques can work wonder in the 

\subsection{NOMA with FD}\label{sec:FullDuplexing}
\textbf{FD}\normalmarginpar\marginnote{FD is a mode of communication wherein the date is transmitted and received at the same time and over the same channel} is a mode of communication wherein the date is transmitted and received at the same time and over the same channel. Conventionally, data transmission and data receiving is done either on different channels or at different time instant or both. However, as a move towards utilizing the spectrum efficiently, the thought of transmitting and receiving the data at the same time and over the same channel is seen as a candidate approach which is technically termed as FD \cite{full_duplex_sabharwal}. FD enables reliable uplink and downlink transmissions simultaneously. However, this could cause interference, which is termed as self interference (SI), as shown in Fig.~\ref{fig:2} (IX). As can be inferred from the application of NOMA and full duplexing, they both are complementary in principle. Hence, merging the two techniques can prove fruitful and play potential role in improving the spectral efficiency of 5G networks \cite{full_duplex_NOMA_MC}. Furthermore, as discussed in Section~\ref{sec:SWIPT}, the cooperation in NOMA network can be performed at the same time instant by using FD such that cell center user can receive the data from the base station and also retransmit the data to the cell edge user.
Hence, integrating NOMA with full duplexing can be beneficial in scenarios when full duplex base station served uplink and downlink users at the same time using same channel \cite{full_duplex_NOMA_MC}, when cooperative NOMA is applied using full duplex NOMA cell center users \cite{full_duplex_cooperative_NOMA}, and when relay assisted communication is performed using full duplex relays \cite{full_duplex_relaying_NOMA}. FD increases the number of connected devices at a given instant of time by promoting uplink and downlink transmissions simultaneously. Furthermore, FD integrated with NOMA augments number of uplink and downlink connections.
%The proliferation of communication devices and high-speed data services has led to a demand for increased spectral efficiency in emerging wireless systems such as 5G. To meet this demand,full-duplex communications, that is, simultaneously transmitting and receiving radio signals on the same frequency band, has become a promising solution \cite{full_duplex_sabharwal}. Since NOMA and full-duplex are complementary in principle, their integration is worthwhile to study in detail \cite{full_duplex_NOMA_MC}.A typical objective of full-duplex transceivers in NOMA systems is to enable reliable and simultaneous transmission in uplink (UL) and DL channels. More specifically, data from UL users and data to be sent to DL users are received and transmitted simultaneously at the same frequency. Furthermore, by incorporating the full-duplex operation, a NOMA user close to the base station (BS) can simultaneously receive and forward data to a distant NOMA user. 
%Two cases of the NOMA concept benefting from full-duplex operation are when a full-duplex BS serves UL and DL users at the same time in the same frequency \cite{full_duplex_NOMA_MC}, and cooperative NOMA systems where full-duplex NOMA-strong users \cite{full_duplex_cooperative_NOMA},or dedicated full-duplex relays \cite{full_duplex_relaying_NOMA} assist the transmissions between the source and NOMA-weak users.

\subsection{Hybrid NOMA techniques}\label{hybrid}
\begin{figure}
 \includegraphics[scale=1.0]{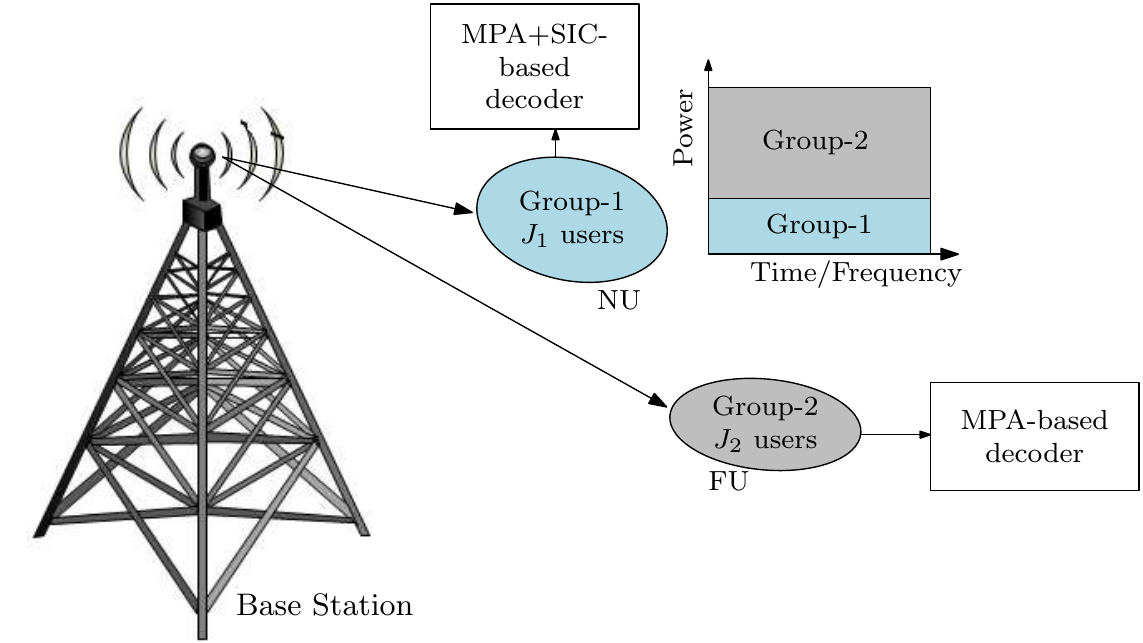}
\caption{Hybrid multiple access method  for 5G and beyond.}
\label{hybrid1}       % Give a unique label
\end{figure}

In a practical system, both cell center users and cell edge users exists. Furthermore, there may exists difference in number of cell center users (assumed less here) and cell edge users (assumed more here) due to a higher area of peripheral than the cell center area of a base station. Therefore,\normalmarginpar\marginnote{Users' channel gain difference and their distribution around base station motivates the use of a hybrid MA (HMA) system by combining the PD-NOMA and SCMA techniques.} users' channel gain difference and their distribution around base station motivates the use of a hybrid MA (HMA) system by combining the PD-NOMA and SCMA techniques. HMA enhances the spectral efficiency and accommodates a large number of users in the system as compared to the conventional PD-NOMA or SCMA for next-generation networks \cite{sanjeev_sharma}. In HMA, both message passing algorithm (MPA) and SIC based method is used for detection \cite{sanjeev_sharma}. For instance, consider a downlink system with one base station and $J$ users, such that $J_1$ users cell center users (called as Group-1) and $J_2$ are cell edge users (called as Group-1), as shown in Fig. \ref{hybrid1}.  Users within a group use the SCMA technique and groups can be distinguished using PD-NOMA techniques. Therefore, users of both the groups can communicate using say orthogonal resources in the system. Therefore, spectral efficiency of HMA-based system is higher than conventional PD-NOMA and SCMA with higher flexibility in terms of user pairing.

\section{Future Challenges}\label{sec:FW}
Although a lot of literature exists in the co-existence of NOMA with other emerging technologies, a few technical challenges that need to be addressed in detail remains to be explored. These challenges include optimal user pairing for NOMA in massive MIMO setups, design challenges of integrating NOMA techniques into UAV networks. Investigation of VLC-MIMO NOMA in detail, eliminating intercell interference in H-UDNs with NOMA, evaluating the robustness of NOMA against synchronization errors and inter-cell interference when applied in a multi-cell scenario, as well as the impact of partial interference cancellation caused by imperfect channel estimation. Furthermore, theoretical performance analysis for a better understanding of the impact of NOMA on MEC needs careful study. However, it is clear and pertinent that NOMA is an enabler for massive connectivity.

%\begin{acknowledgements}
%\end{acknowledgements}

\bibliographystyle{IEEEtran}
\bibliography{arXiv}

\begin{thebibliography}{10}
\providecommand{\url}[1]{#1}
\csname url@samestyle\endcsname
\providecommand{\newblock}{\relax}
\providecommand{\bibinfo}[2]{#2}
\providecommand{\BIBentrySTDinterwordspacing}{\spaceskip=0pt\relax}
\providecommand{\BIBentryALTinterwordstretchfactor}{4}
\providecommand{\BIBentryALTinterwordspacing}{\spaceskip=\fontdimen2\font plus
\BIBentryALTinterwordstretchfactor\fontdimen3\font minus
  \fontdimen4\font\relax}
\providecommand{\BIBforeignlanguage}[2]{{%
\expandafter\ifx\csname l@#1\endcsname\relax
\typeout{** WARNING: IEEEtran.bst: No hyphenation pattern has been}%
\typeout{** loaded for the language `#1'. Using the pattern for}%
\typeout{** the default language instead.}%
\else
\language=\csname l@#1\endcsname
\fi
#2}}
\providecommand{\BIBdecl}{\relax}
\BIBdecl

\bibitem{KPI_5G}
M.~Series, ``{IMT Vision--Framework and overall objectives of the future
  development of {IMT} for 2020 and beyond},'' \emph{Recommendation ITU}, pp.
  2083--0, 2015.

\bibitem{impactofuserpairing}
Z.~Ding, P.~Fan, and H.~V. Poor, ``Impact of user pairing on {5G} nonorthogonal
  multiple-access downlink transmissions,'' \emph{IEEE Trans. Veh. Technol.},
  vol.~65, no.~8, pp. 6010--6023, 2016.

\bibitem{IDMA}
L.~Ping, L.~Liu, K.~Wu, and W.~K. Leung, ``Interleave division
  multiple-access,'' \emph{IEEE Trans. Wireless Commun.}, vol.~5, no.~4, pp.
  938--947, 2006.

\bibitem{PD_NOMA}
K.~Higuchi and A.~Benjebbour, ``Non-orthogonal multiple access {(NOMA)} with
  successive interference cancellation for future radio access,'' \emph{IEICE
  Trans. Commun.}, vol.~98, no.~3, pp. 403--414, 2015.

\bibitem{MUST}
3GPP, Study on Downlink Multiuser Superposition Transmission (MUST) for LTE
  (Release 13), TR36.859, Dec. 2015.

\bibitem{3GPP_NTT}
3GPP, NTT-DOCOMO, Initial views and evaluation results on non-orthogonal
  multiple access for NR, R1-165175, May 2016.

\bibitem{low_density_CDMA}
R.~Hoshyar, F.~P. Wathan, and R.~Tafazolli, ``Novel low-density signature for
  synchronous {CDMA} systems over {AWGN} channel,'' \emph{IEEE Trans. Signal
  Process.}, vol.~56, no.~4, pp. 1616--1626, 2008.

\bibitem{SCMA}
3GPP, Huawei, HiSilicon, Sparse Code Multiple Access (SCMA) for 5G Radio
  Transmission, R1- 162155, Apr. 2016.

\bibitem{PDMA}
X.~Dai, Z.~Zhang, B.~Bai, S.~Chen, and S.~Sun, ``Pattern division multiple
  access: A new multiple access technology for {5G},'' \emph{IEEE Wireless
  Commun.}, vol.~25, no.~2, pp. 54--60, 2018.

\bibitem{MUSA}
Z.~Yuan, G.~Yu, W.~Li, Y.~Yuan, X.~Wang, and J.~Xu, ``Multi-user shared access
  for internet of things,'' in \emph{IEEE Vehicular Technology Conference (VTC
  Spring)}, 2016, pp. 1--5.

\bibitem{84BIS}
3GPP, MediaTek Inc., CMCC, etc. New work item proposal: Downlink Multiuser
  Superposition Transmission for LTE, RP-160680, Mar. 2016.

\bibitem{VTC}
P.~Swami, V.~Bhatia, S.~Vuppala, and T.~Ratnarajah, ``User fairness and
  performance enhancement for cell edge user in noma-hcn with offloading,'' in
  \emph{IEEE Vehicular Technology Conference (VTC Spring)}.\hskip 1em plus
  0.5em minus 0.4em\relax IEEE, 2017, pp. 1--5.

\bibitem{GLOBECOM_optimization}
------, ``{Joint Optimization of Power Allocation and Channel Ratio for
  Offloading in NOMA-HetNets},'' in \emph{IEEE Globecom Workshops (GC
  Wkshps)}.\hskip 1em plus 0.5em minus 0.4em\relax IEEE, 2018, pp. 1--6.

\bibitem{saito}
R.~Razavi, M.~Dianati, and M.~A. Imran, ``Non-orthogonal multiple access
  ({NOMA}) for future radio access,'' in \emph{{5G} Mobile Commun.}\hskip 1em
  plus 0.5em minus 0.4em\relax Springer, 2017, pp. 135--163.

\bibitem{recommendation_ITU_5G}
Recommendation {ITU-R M.2083}: {IMT Vision}, Framework and overall objectives
  of the future development of IMT for 2020 and beyond, Sep. 2015.

\bibitem{multipleaccessforUL}
3GPP, Huawei, HiSilicon, Multiple access for UL small packets transmission,
  R1-164036, May 2016.

\bibitem{usecases_5G}
3GPP, CATT, Discussion on scenarios and use cases for multiple access,
  R1-164246, May.2016.

\bibitem{chandra_mag}
K.~Chandra, A.~S. Marcano, S.~Mumtaz, R.~V. Prasad, and H.~L. Christiansen,
  ``Unveiling capacity gains in ultradense networks: Using mm-{W}ave {NOMA},''
  \emph{IEEE Veh. Technol. Mag.}, vol.~13, no.~2, pp. 75--83, 2018.

\bibitem{NOMAmmW7}
L.~Zhu, J.~Zhang, Z.~Xiao, X.~Cao, D.~O. Wu, and X.-G. Xia, ``Joint {Tx-Rx}
  beamforming and power allocation for {5G} millimeter-wave non-orthogonal
  multiple access (mmwave-noma) networks,'' \emph{IEEE Trans. Commun.}, 2019.

\bibitem{mmW_outdoor}
A.~Thornburg, T.~Bai, and R.~W. Heath~Jr, ``Performance analysis of outdoor
  {mmWave} ad hoc networks.'' \emph{IEEE Trans. Signal Process.}, vol.~64,
  no.~15, pp. 4065--4079, 2016.

\bibitem{initial_beamtraining}
Y.~Li, J.~G. Andrews, F.~Baccelli, T.~D. Novlan, and J.~Zhang, ``On the initial
  access design in millimeter wave cellular networks,'' in \emph{Globecom
  Workshops (GC Wkshps), 2016 IEEE}.\hskip 1em plus 0.5em minus 0.4em\relax
  IEEE, 2016, pp. 1--6.

\bibitem{RAT}
G.~Ghatak, A.~De~Domenico, and M.~Coupechoux, ``Coverage analysis and load
  balancing in {HetNets} with millimeter wave multi-{RAT} small cells,''
  \emph{IEEE Trans. Wireless Commun.}, vol.~17, no.~5, pp. 3154--3169, 2018.

\bibitem{ISWCS}
P.~Swami, M.~K. Mishra, V.~Bhatia, and T.~Ratnarajah, ``{Outage Probability of
  Ultra High Frequency and Millimeter Wave Based HetNets with NOMA},'' in
  \emph{16th Int. Symposium Wireless Commun. Syst. (ISWCS)}.\hskip 1em plus
  0.5em minus 0.4em\relax IEEE, 2019, pp. 166--170.

\bibitem{UDN_mag}
Z.~Zhang, G.~Yang, Z.~Ma, M.~Xiao, Z.~Ding, and P.~Fan, ``Heterogeneous
  ultradense networks with {NOMA}: System architecture, coordination framework,
  and performance evaluation,'' \emph{IEEE Veh. Technol. Mag.}, vol.~13, no.~2,
  pp. 110--120, 2018.

\bibitem{UDN_NOMA_Mag}
Z.~Qin, X.~Yue, Y.~Liu, Z.~Ding, and A.~Nallanathan, ``User association and
  resource allocation in unified {NOMA} enabled heterogeneous ultra dense
  networks,'' \emph{IEEE Commun. Mag.}, vol.~56, no.~6, pp. 86--92, 2018.

\bibitem{MY_TVT}
P.~Swami, V.~Bhatia, S.~Vuppala, and T.~Ratnarajah, ``A cooperation scheme for
  user fairness and performance enhancement in {NOMA-HCN},'' \emph{IEEE Trans.
  Veh. Technol.}, vol.~67, no.~12, pp. 11\,965--11\,978, 2018.

\bibitem{MY_ISJ}
------, ``On user offloading in {NOMA-HetNet} using repulsive point process,''
  \emph{IEEE Syst. J.}, vol.~13, no.~2, pp. 1409--1420, 2018.

\bibitem{network_densification}
V.~M. Nguyen and M.~Kountouris, ``Performance limits of network
  densification,'' \emph{arXiv preprint arXiv:1611.07790}, 2016.

\bibitem{architecture_5G}
{System Architecture for the 5G System, 3GPP TS 23.501, 2017}.

\bibitem{SDN}
F.~Granelli, A.~A. Gebremariam, M.~Usman, F.~Cugini, V.~Stamati, M.~Alitska,
  and P.~Chatzimisios, ``Software defined and virtualized wireless access in
  future wireless networks: scenarios and standards,'' \emph{IEEE Commun.
  Mag.}, vol.~53, no.~6, pp. 26--34, 2015.

\bibitem{cloud_computing}
A.~Checko, H.~L. Christiansen, Y.~Yan, L.~Scolari, G.~Kardaras, M.~S. Berger,
  and L.~Dittmann, ``{Cloud RAN for mobile networks-A technology overview},''
  \emph{IEEE Commun. surveys \& tutorials}, vol.~17, no.~1, pp. 405--426, 2014.

\bibitem{fog_computing}
M.~Peng, S.~Yan, K.~Zhang, and C.~Wang, ``Fog-computing-based radio access
  networks: Issues and challenges,'' \emph{IEEE Network}, vol.~30, no.~4, pp.
  46--53, 2016.

\bibitem{AF_DF_CR}
D.~Li, ``{Opportunistic DF-AF selection for cognitive relay networks},''
  \emph{IEEE Trans. Veh. Technol.}, vol.~65, no.~4, pp. 2790--2796, 2015.

\bibitem{AF_DF_NOMA}
Y.~Liu, G.~Pan, H.~Zhang, and M.~Song, ``{Hybrid decode-forward \&
  amplify-forward relaying with non-orthogonal multiple access},'' \emph{IEEE
  Access}, vol.~4, pp. 4912--4921, 2016.

\bibitem{full_duplex_cooperative_NOMA}
Z.~Zhang, Z.~Ma, M.~Xiao, Z.~Ding, and P.~Fan, ``Full-duplex
  device-to-device-aided cooperative nonorthogonal multiple access,''
  \emph{IEEE Trans. Veh. Technol.}, vol.~66, no.~5, pp. 4467--4471, 2016.

\bibitem{ICACCI}
P.~Swami, M.~K. Mishra, and A.~Trivedi, ``Performance analysis of two-tier
  cellular network using power control and cooperation,'' in
  \emph{International Conf. Advances Comput., Commun. Informatics
  ({ICACCI})}.\hskip 1em plus 0.5em minus 0.4em\relax IEEE, 2016, pp. 322--327.

\bibitem{IJCS}
\BIBentryALTinterwordspacing
------, ``Analysis of downlink power control and cooperation scheme for
  two-tier heterogeneous cellular network,'' \emph{International J. Commun.
  Syst.}, pp. e3282--n/a, 2017, e3282 dac.3282. [Online]. Available:
  \url{http://dx.doi.org/10.1002/dac.3282}
\BIBentrySTDinterwordspacing

\bibitem{SWIPT_Firstpaper}
L.~R. Varshney, ``Transporting information and energy simultaneously,'' in
  \emph{IEEE Int. Symposium Info. Theory}.\hskip 1em plus 0.5em minus
  0.4em\relax IEEE, 2008, pp. 1612--1616.

\bibitem{cooperative_NOMA_Ding}
Z.~Ding, M.~Peng, and H.~V. Poor, ``Cooperative non-orthogonal multiple access
  in {5G} systems,'' \emph{IEEE Commun. Lett.}, vol.~19, no.~8, pp. 1462--1465,
  2015.

\bibitem{VLC_Mag}
M.~Figueiredo, L.~N. Alves, and C.~Ribeiro, ``{Lighting the wireless world: The
  promise and challenges of visible light communication},'' \emph{IEEE Consum.
  Electron. Mag.}, vol.~6, no.~4, pp. 28--37, 2017.

\bibitem{NOMA_VLC}
L.~Yin, W.~O. Popoola, X.~Wu, and H.~Haas, ``Performance evaluation of
  non-orthogonal multiple access in visible light communication,'' \emph{IEEE
  Trans. Commun.}, vol.~64, no.~12, pp. 5162--5175, 2016.

\bibitem{LED_nonlinearity}
B.~Inan, S.~J. Lee, S.~Randel, I.~Neokosmidis, A.~M. Koonen, and J.~W.
  Walewski, ``{Impact of LED nonlinearity on discrete multitone modulation},''
  \emph{J. Opt. Commun. Netw.}, vol.~1, no.~5, pp. 439--451, 2009.

\bibitem{NOMA_VLC2}
H.~Marshoud, V.~M. Kapinas, G.~K. Karagiannidis, and S.~Muhaidat,
  ``Non-orthogonal multiple access for visible light communications,''
  \emph{IEEE Photon. Technol. Lett.}, vol.~28, no.~1, pp. 51--54, 2015.

\bibitem{double_heterostructure_LED}
T.~Lee, ``{The nonlinearity of double-heterostructure LED's for optical
  communications},'' \emph{Proceedings of the IEEE}, vol.~65, no.~9, pp.
  1408--1410, 1977.

\bibitem{mitra}
R.~Mitra and V.~Bhatia, ``Precoded {C}hebyshev-{NLMS} based pre-distorter for
  nonlinear {LED} compensation in {NOMA-VLC},'' \emph{IEEE Trans. Commun.},
  2017.

\bibitem{Massive_mimo_mag}
E.~G. Larsson, O.~Edfors, F.~Tufvesson, and T.~L. Marzetta, ``Massive {MIMO}
  for next generation wireless systems,'' \emph{IEEE Commun. Mag.}, vol.~52,
  no.~2, pp. 186--195, 2014.

\bibitem{MIMO_mobile_broadband}
G.~Liu, X.~Hou, J.~Jin, F.~Wang, Q.~Wang, Y.~Hao, Y.~Huang, X.~Wang, X.~Xiao,
  and A.~Deng, ``{3-D-MIMO} with massive antennas paves the way to {5G}
  enhanced mobile broadband: {F}rom system design to field trials,'' \emph{IEEE
  J. Sel. Areas Commun.}, vol.~35, no.~6, pp. 1222--1233, 2017.

\bibitem{unlimitedantennas}
T.~L. Marzetta, ``Noncooperative cellular wireless with unlimited numbers of
  base station antennas,'' \emph{IEEE Trans. Wireless Commun.}, vol.~9, no.~11,
  pp. 3590--3600, 2010.

\bibitem{NOMA_applcation_to_MIMO}
Z.~Ding, F.~Adachi, and H.~V. Poor, ``{The application of MIMO to
  non-orthogonal multiple access},'' \emph{IEEE Trans. Wireless Commun.},
  vol.~15, no.~1, pp. 537--552, 2015.

\bibitem{MIMO_NOMA}
Z.~Ding, R.~Schober, and H.~V. Poor, ``A general {MIMO} framework for {NOMA}
  downlink and uplink transmission based on signal alignment,'' \emph{IEEE
  Trans. Wireless Commun.}, vol.~15, no.~6, pp. 4438--4454, 2016.

\bibitem{NOMAroleinMassiveMIMO}
K.~Senel, H.~V. Cheng, E.~Bj{\"o}rnson, and E.~G. Larsson, ``{What role can
  NOMA play in massive MIMO?}'' \emph{IEEE J. Sel. Topics Signal Process.},
  vol.~13, no.~3, pp. 597--611, 2019.

\bibitem{underlay_overlay}
M.~G. Khoshkholgh, K.~Navaie, and H.~Yanikomeroglu, ``{Access strategies for
  spectrum sharing in fading environment: Overlay, underlay, and mixed},''
  \emph{IEEE Trans. Mobile Comput.}, vol.~9, no.~12, pp. 1780--1793, 2010.

\bibitem{NOMA_CR1}
Z.~Song, X.~Wang, Y.~Liu, and Z.~Zhang, ``{Joint spectrum resource allocation
  in NOMA-based cognitive radio network with SWIPT},'' \emph{IEEE Access},
  vol.~7, pp. 89\,594--89\,603, 2019.

\bibitem{NOMA_CR2}
W.~Xu, R.~Qiu, and X.-Q. Jiang, ``{Resource Allocation in Heterogeneous
  Cognitive Radio Network With Non-Orthogonal Multiple Access},'' \emph{IEEE
  Access}, vol.~7, pp. 57\,488--57\,499, 2019.

\bibitem{virtual_reality}
E.~Bastug, M.~Bennis, M.~M{\'e}dard, and M.~Debbah, ``Toward interconnected
  virtual reality: Opportunities, challenges, and enablers,'' \emph{IEEE
  Commun. Mag.}, vol.~55, no.~6, pp. 110--117, 2017.

\bibitem{NOMA_MEC1}
F.~Wang, J.~Xu, and Z.~Ding, ``Optimized multiuser computation offloading with
  multi-antenna {NOMA},'' in \emph{IEEE Globecom Workshops}, 2017, pp. 1--7.

\bibitem{NOMA_MEC2}
A.~Kiani and N.~Ansari, ``Edge computing aware {NOMA} for {5G} networks,''
  \emph{IEEE Internet Things J.}, vol.~5, no.~2, pp. 1299--1306, 2018.

\bibitem{airborne_comm}
E.~W. Frew and T.~X. Brown, ``Airborne communication networks for small
  unmanned aircraft systems,'' \emph{Proceedings of the {IEEE}}, vol.~96,
  no.~12, 2008.

\bibitem{facebook_drone}
M.~Mozaffari, W.~Saad, M.~Bennis, and M.~Debbah, ``Wireless communication using
  unmanned aerial vehicles {(UAVs)}: {O}ptimal transport theory for hover time
  optimization,'' \emph{IEEE Trans. Wireless Commun.}, vol.~16, no.~12, pp.
  8052--8066, 2017.

\bibitem{UAV_NOMA}
Y.~Liu, Z.~Qin, Y.~Cai, Y.~Gao, G.~Y. Li, and A.~Nallanathan, ``Uav
  communications based on non-orthogonal multiple access,'' \emph{IEEE Wireless
  Commun.}, vol.~26, no.~1, pp. 52--57, 2019.

\bibitem{full_duplex_sabharwal}
A.~Sabharwal, P.~Schniter, D.~Guo, D.~W. Bliss, S.~Rangarajan, and R.~Wichman,
  ``In-band full-duplex wireless: {C}hallenges and opportunities,'' \emph{IEEE
  J. Sel. Areas Commun.}, vol.~32, no.~9, pp. 1637--1652, 2014.

\bibitem{full_duplex_NOMA_MC}
Y.~Sun, D.~W.~K. Ng, Z.~Ding, and R.~Schober, ``Optimal joint power and
  subcarrier allocation for full-duplex multicarrier non-orthogonal multiple
  access systems,'' \emph{IEEE Trans. Commun.}, vol.~65, no.~3, pp. 1077--1091,
  2017.

\bibitem{full_duplex_relaying_NOMA}
C.~Zhong and Z.~Zhang, ``Non-orthogonal multiple access with cooperative
  full-duplex relaying,'' \emph{IEEE Commun. Lett.}, vol.~20, no.~12, pp.
  2478--2481, 2016.

\bibitem{sanjeev_sharma}
S.~Sharma, K.~Deka, V.~Bhatia, and A.~Gupta, ``Joint power-domain and
  {SCMA}-based {NOMA} system for downlink in 5{G} and beyond,'' \emph{IEEE
  Communications Letters}, vol.~23, no.~6, pp. 971--974, 2019.

\end{thebibliography}
\vskip 0pt plus -1fil
\begin{IEEEbiography}[{\includegraphics[width=1.5in,height=1.25in,clip,keepaspectratio]{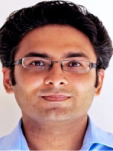}}]%
{Vimal Bhatia}
(M'99-SM'12) received the Ph.D.degree from the Institute for Digital Communications, The University of Edinburgh, U.K., in 2005.During the work for the Ph.D. degree he received the IEE fellowship for collaborative research on OFDM with Prof. Falconer at Carleton University,Canada. He has been with various telecommunication and new media companies in the U.K. and India. He is currently an Associate Professor with IIT Indore, India. He is the author or co-author of around 100 technical papers in scientific journals and presented at international conferences. He also holds nine filed and two granted patents. As co-author, he has received the best paper awards at OSA and IEEE conferences, and was awarded the Young Faculty Research Fellow Award from the Ministry of Electronics and Information Technology, India. His research interests are in 5G networks, visible light communication, MIMO systems, and statistical signal processing. He is serving as an Editor of the IETE Technical Review and is reviewer for IEEE, OSA, Elsevier, and Springer. He is also a certified SCRUM Master.
\end{IEEEbiography}
\vskip 0pt plus -1fil
\begin{IEEEbiography}[{\includegraphics[width=1in,height=1.25in,clip,keepaspectratio]{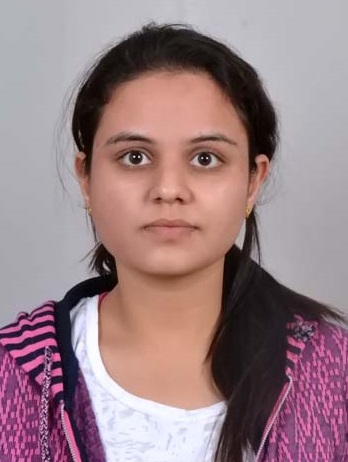}}]%
{Pragya Swami}
received the B.Tech degree from the Indian Institute of Information Technology, Design and Manufacturing, Jabalpur, India, in 2014, the M.Tech degree from the Indian Institute of Information Technology and Management Gwalior, India, in 2016. Currently, she is pursuing Ph.D. from the Indian Institute of Technology Indore, India. Her research interests include performance analysis of 5G and beyond networks using non-orthogonal multiple access.
\end{IEEEbiography}
\vskip 0pt plus -1fil
\begin{IEEEbiography}[{\includegraphics[width=1in,height=1.25in,clip,keepaspectratio]{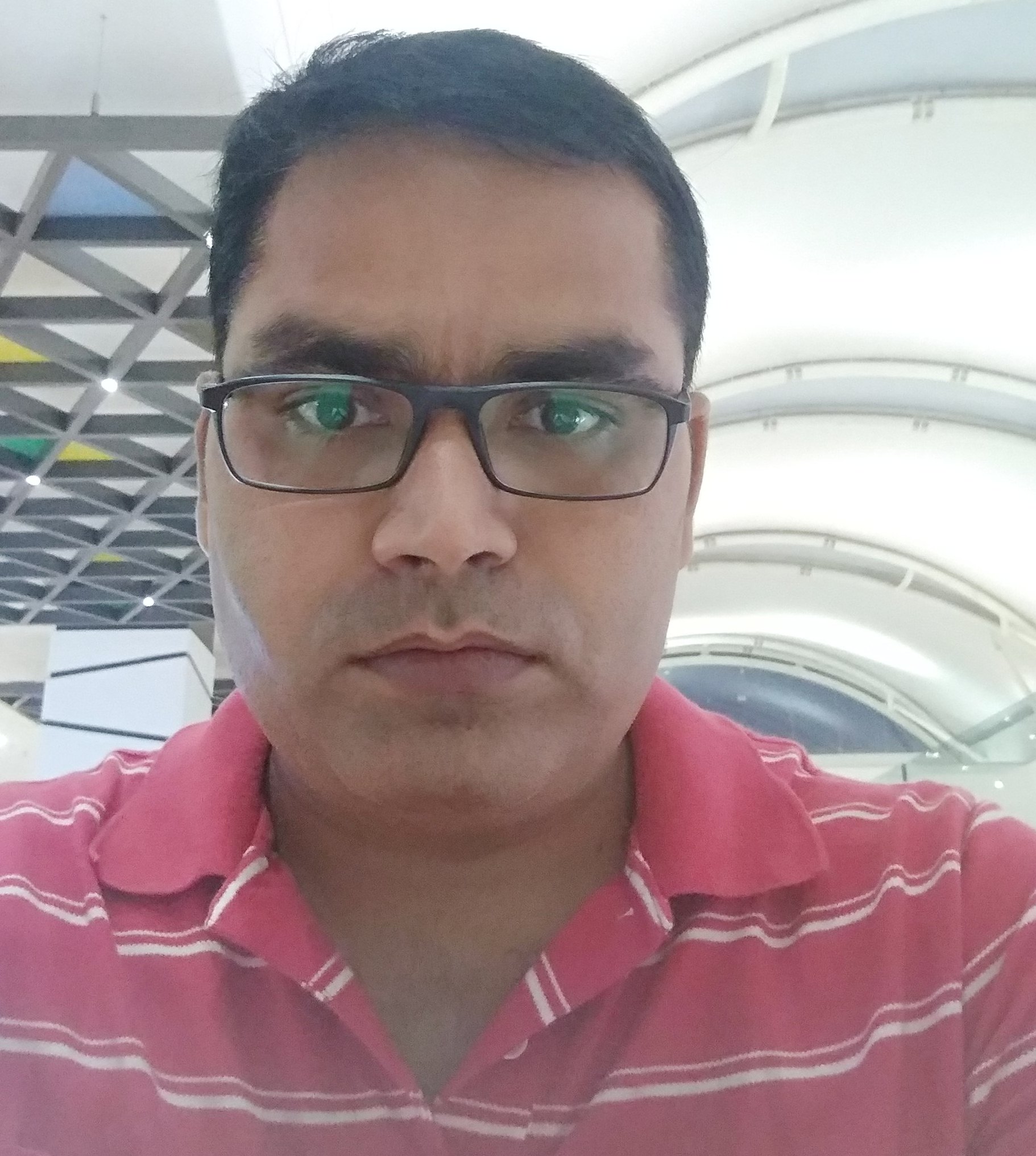}}]%
{Sanjeev Sharma}
received the M.Tech. degree from the Indian Institute of Technology Guwahati, Guwahati, India, in 2010 and the Ph.D. degree from the  Indian Institute of Technology Indore, Indore, India, in 2018. He is currently an Assistant Professor  in the Department of Electronics Engineering at IIT (BHU) Varanasi, India. His main research interests include robust receiver design and sparsity-based signal processing for next-generation wireless communications.
\end{IEEEbiography}\vskip 0pt plus -1fil
\begin{IEEEbiography}[{\includegraphics[width=1in,height=1.25in,clip,keepaspectratio]{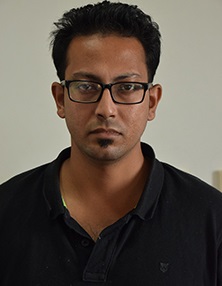}}]%
{Rangeet Mitra}
received the B.Tech. degree from Asansol Engineering College, WBUT, India, in 2008, the M.Tech. degree from Indian Institute of Technology Guwahati, India, in 2010, and the Ph.D. degree from Indian Institute of Technology Indore, India, in 2017. His research interests include adaptive signal processing in RKHS and nonlinear signal processing for upcoming 5G-based communication systems.
\end{IEEEbiography}
%% insert where needed to balance the two columns on the last page with
%%\newpage


%\vfill


\end{document}